# The Impact of Employee Education and Health on Firm-Level TFP in China

He Yuhan (ID:He111OSRP24)

25/7/2024

**Abstract**

This study examines the influence of employee education and health on firm-level Total Factor Productivity (TFP) in China, using panel data from A-share listed companies spanning from 2007 to 2022. The analysis shows that life expectancy and higher education have a significant impact on TFP. More optimal health conditions can result in increased productivity through decreased absenteeism and improved work efficiency. Similarly, higher levels of education can support technological adaptation, innovation, and managerial efficiency. Nevertheless, the correlation between health and higher education indicates that there may be a point where further improvements in health yield diminishing returns in terms of productivity for individuals with advanced education. These findings emphasise the importance of implementing comprehensive policies that improve both health and education, maximising their impact on productivity. This study adds to the current body of research by presenting empirical evidence at the firm-level in China. It also provides practical insights for policymakers and business leaders who want to improve economic growth and competitiveness. Future research should take into account wider datasets, more extensive health metrics, and delve into the mechanisms that contribute to the diminishing returns observed in the relationship between health and education.

**Introduction**

TFP is an essential measure for evaluating firm productivity. It demonstrates the growth in output that goes beyond the usual factors of capital and labour, taking into account advancements in technology, efficiency in management, and other intangible elements (Solow, 1956). In recent years, there has been a growing interest in TFP due to the rapid pace of globalisation and technological advancements. It is widely acknowledged in academic circles that the educational background and health of employees play a crucial role in determining the firm-level TFP (Alvi & Ahmed, 2014; Bloom et al., 1999; Knowles & Owen, 1995). Past research has indicated that education and health have a positive impact on both labour productivity and the ability of firms to innovate and compete. This has been demonstrated through various channels in studies conducted by Barro & Sala-i-Martin (2004) and Aghion & Howitt (1998). Additionally, as a Chinese resident, I have observed significant advancements in education and healthcare among the Chinese population, which undoubtedly have an impact on firm-level TFP. China's policies on education and health have proven to be effective in recent years. China's education system has undergone significant improvements since the amendment of the Compulsory Education Law in 2006. With the implementation of various policies to safeguard education, the average number of years of education for the population has increased from 7.1 years in 2000 to 9.7 years in 2022, leading to a notable enhancement in the quality of the population. Furthermore, China has experienced notable improvements in its overall health since the implementation of the Healthy China 2030 plan in 2016. In the year 2000, China had an average life expectancy of 71.4 years. In just over two decades, the average life expectancy has increased significantly, reaching 78.6 years by 2022. This remarkable progress has propelled its global ranking from 96th in 2000 to an impressive 53rd in 2022. (World Bank, 2022) However, there is a lack of extensive research on firm-level data in China, especially studies that analyse the combined effects of education and health on TFP.

With the rapid growth of China's economy in recent years, studying TFP in Chinese firms has gained significant importance. Previous studies have shown that improving TFP is crucial for long-term economic growth. Additionally, education and health, which are important aspects of human capital, have substantial effects on TFP (Barro & Lee, 2010; Alvi & Ahmed, 2014).

In this study, the measurement of education level is based on the percentage of employees with different educational achievements. Education is widely recognised as a crucial factor in enhancing worker skills and productivity (Barro & Lee, 2010). Higher levels of education not only improve employees' professional skills, but also enhance their ability to learn and their awareness of innovation, which in turn boosts overall productivity within a company. Life expectancy is a measure of health status, as it is believed that good health enhances worker efficiency and productivity (WHO, 2008). Having a healthy workforce can significantly decrease the number of sick days taken and improve overall energy and concentration levels, resulting in a boost in productivity.

**Literature Review**

Previous research can be broadly categorised into the following groups. One area of focus is the influence of education on TFP. According to research conducted by Mankiw et al. (1992), it is widely believed that education plays a crucial role in enhancing worker skills and knowledge, ultimately leading to an improvement in TFP. There are two main perspectives that can be used to further categorise this topic. According to one perspective, higher education has a profound influence on TFP by greatly improving innovation capabilities and the implementation of technology (Lucas, 1988). This viewpoint highlights the importance of employees with advanced education in their ability to adapt to modern production technologies and management practices, which has a positive impact on firm TFP. Another perspective suggests that basic education is just as crucial as it provides the groundwork for advanced education and vocational training (Romer, 1990). This viewpoint suggests that a strong foundation in primary education is necessary for successful tertiary education and vocational training, both of which play a crucial role in enhancing the efficiency of workers. It is widely agreed upon in these studies that education has a significant impact on TFP.

Another area of study focusses on how health affects TFP. The prevailing perspective in this field suggests that the state of one's health plays a vital role in enhancing labour productivity and total factor productivity (Bloom et al., 1999). This category can be further divided into two representative views. According to a certain perspective, there is a claim that good health has a direct impact on TFP by decreasing absenteeism and enhancing work efficiency (Gallup & Sachs, 2000). This viewpoint emphasises the positive impact of employee health on workplace efficiency and productivity, which in turn contributes to overall firm productivity. According to another perspective, the improvement of employees' learning and training abilities can have an indirect impact on TFP by means of better health (Mayer, 2001). This viewpoint suggests that employees who are in good health are better equipped to engage in training and education, which in turn enhances their work efficiency and productivity. While these studies suggest that health has a positive effect on TFP, the majority of them primarily examine the broader macroeconomic perspective, with less emphasis on analysing firm-level data.

In addition, certain studies (e.g., Alvi & Ahmed, 2014) have examined the combined influence of education and health on TFP from a holistic standpoint. These studies indicate that education and health have a symbiotic relationship, where a good education amplifies the advantages of good health, and vice versa. Nevertheless, there remains a lack of research in this area, specifically in terms of empirical studies at the firm level.

More specifically, there are certain limitations that can be identified in the existing research. First, previous research primarily concentrates on macro-level data, with limited utilisation of firm-level microdata. This study examines the effects of education and health on TFP using firm-level data from China, providing a detailed analysis from a micro perspective (Barro & Lee, 2010; Alvi & Ahmed, 2014). Utilising firm-level data provides a more precise examination of the direct effects of education and health on firm productivity. Additionally, many previous studies have relied on cross-sectional data and have not fully utilised the benefits of panel data analysis methods. This study utilises panel data from 2007-2022, providing a more precise depiction of dynamic changes over time (Alvi & Ahmed, 2014). By utilising a broader range of data over an extended period, one can more effectively analyse the lasting effects of education and health on TFP. In addition, previous studies tend to focus on analysing the individual effects of education or health on TFP, without fully considering their combined impact. This study investigates the influence of education and health on TFP and explores how they work together (Lucas, 1988; Romer, 1990). Through a thorough analysis of the effects of education and health, a more comprehensive understanding of their role in TFP can be attained.

This study aims to provide new perspectives on understanding the impact of education and health on firm productivity and offer valuable insights for policymakers. This research not only addresses gaps in the existing literature, but also offers empirical evidence to enhance TFP in Chinese firms.

**Theoretical Framework and Hypothesis**

The relationship between health and education and their impact on TFP has been widely debated for many years. The foundational models developed by Becker (1964) and Ben-Porath (1967) sought to identify the most beneficial investments in human capital. Grossman's Human Capital Model (1972) made significant advancements in the field of health economics by analysing health demand through the perspective of human capital theory. According to Grossman's model, better health, reflected in longer life expectancy, directly boosts productivity by enhancing the availability and efficiency of the workforce. At the same time, higher levels of education also have a significant impact on a society's ability to acquire and apply relevant knowledge, thereby greatly affecting TFP by encouraging technology transfer, innovation, and the increase of employees' learning ability (Mincer, 1974; Romer, 1990; Bartel, 1992). This study argues that health and education have a significant impact on improving TFP at the organisational level. The following discussion provides further details on this topic.

**Health plays a significant role in improving TFP by reducing presenteeism and fostering creative job crafting.** First, **improving health can enhance working efficiency by reducing presenteeism.** Workers who face health challenges or have low expectations for their well-being may struggle with presenteeism, a situation where they are physically present at work but are unable to perform at their optimal level due to health issues. Research has shown a clear link between increased health risks and productivity losses, as well as a correlation between lower health risks and reduced production losses (Burton, W. N., et al, 2006; Hertz, R. P., et al, 2005) (Wolf Kirsten, 2009). In addition, a study conducted by vielife (2007) provided support for this viewpoint by implementing a health enhancement initiative in a company, resulting in an 8.3% increase in productivity and a return on investment (ROI) of 3.73. Furthermore, **prioritising health can foster a work environment that promotes innovation and ultimately boosts productivity** (Philip & Anrea, 2018). Philip and Anrea (2018) demonstrated that employees are more inclined to engage in promotion-focused job crafting when they are in a better state of well-being. In this situation, individuals proactively improve their work by increasing job resources such as autonomy and social support, while also taking on challenging job demands. This results in enhanced efficiency and production (Philip & Andrea, 2018). On the other hand, better health can also reduce the inclination of employees to engage in prevention-focused job crafting. This type of job crafting involves avoiding demanding tasks and displaying passive behaviour (Philip & Andrea, 2018). Thus, this paper puts forth the following hypothesis.

**H1: Employee's better health condition can improve firm-level TFP**

**Education has a significant impact on TFP by facilitating technology transfer, encouraging innovation, and improving labour learning ability.** First, **education can enhance technology transfer by providing training to employees.** Individuals with a strong educational background are better equipped to implement fresh ideas, adapt to technological advancements, and swiftly introduce new production methods. Ultimately, this results in an enhancement of TFP (Aiyer & Feyrer, 2002; Vandenbussche et al., 2006; Aggrey et al., 2010; Nelson & Phelps, 1966). Furthermore, **education is a vital component of human capital and has a profound impact on driving technological progress and improving efficiency through the creation of positive externalities.** A high-quality education system is crucial for developing a workforce that can drive economic growth through both innovation and imitation. This has been supported by various studies (Nelson & Phelps, 1966; Romer, 1990; Aghion & Howitt, 1998; Benhabib & Spiegel, 1994). Furthermore, **education and training enhance the calibre of the workforce by providing workers with enhanced knowledge, expertise, and understanding.** Acquiring knowledge and expertise through practical experience, commonly known as the "learning-by-doing" impact, enhances productivity. Individuals with a strong educational background possess the

skills to effectively allocate resources, implement innovative technologies, and improve overall productivity (Corvers, 1975; Hao & Wei, 2009; Rasha Qutb, 2017).

On the other hand, individuals with a college-level education contribute significantly to technological progress by conducting research and development (R&D) on new technologies. In contrast, those with a secondary or primary school education may not have the same impact. Romer (1990) highlights the significance of having an R&D sector in an economy as a crucial mechanism for maintaining growth. The level of production is directly influenced by the quality of R&D. Individuals with a high level of education are more inclined to generate innovative ideas, develop new technologies and products, and readily adopt and apply technologies already established by developed nations. University education plays a crucial role in promoting technological progress and diffusion through healthy competition. This leads to positive externalities in terms of innovation and imitation, which can contribute to overall economic growth by altering the diminishing returns of production factors. Therefore, it is anticipated that higher education levels would likely contribute to technological progress. However, the impact of primary and secondary education on technological progress may not be as pronounced. Once technology is acquired, education plays a crucial role in determining how efficiently the new technology is adopted and imitated. If education has a positive impact on efficiency, then the level of education plays a crucial role in determining the technology absorptive capacity of technology users. University-educated individuals have the potential to effectively manage and enhance the efficiency of technology adoption, unlike those with only primary and secondary education. **Thus, it is assumed in this paper that individuals with a university education are more likely to exhibit higher levels of efficiency compared to those with primary and secondary education.**

**H2.1: College and above level education have a positive effect on firm-level TFP**

**H2.2: High School and below level education have no significant effect on**

**firm-level TFP**

While it is true that enhanced health and education conditions can individually contribute to an increase in firm-level TFP, it is important to note that the interaction between health and higher education can potentially have a dynamic effect. **One way to understand the presence of a turning point or threshold in this interaction is by considering a nonlinear relationship, like an inverted U-shape.** It appears that there is a positive relationship between improvements in health and higher education, which can lead to increased productivity. However, after reaching this point, the advantages begin to diminish, like the circumstance of China. The presence of this threshold depends on various factors, including the job's nature, the level of education and health, and the workforce's ability to handle stress and stay productive in challenging situations. For example, in situations where job demands are exceptionally high, the stress and pressure on individuals with advanced education can outweigh the advantages of better health, resulting in decreased marginal productivity (Eklund & Pettersson, 2019; Card et al., 2018). **This paper will examine the situation in China, where the convergence of good health and advanced education tends to have a negative effect on firm-level TFP for several reasons.**

**First, the relationship between health and higher education can cause heightened stress and burnout in individuals with advanced degrees.** This is because they face the challenge of maintaining optimal health standards while already operating at high efficiency levels. As a result, their productivity may diminish and potentially have negative effects. Studies suggest that while better health typically boosts work efficiency by reducing absences and increasing concentration, the relationship between health and advanced education can lead to diminishing returns and potentially adverse outcomes in certain situations. (OECD, 2006) Based on the empirical evidence, it appears that the relationship between health and college education is not simply additive. The negative and significant interaction term implies that the combined effect of these factors is more complex. Individuals with advanced degrees may experience increased levels of stress due to the demands of their

professional responsibilities, which can impact their overall well-being. This excessive stress and overwhelming workload may result in burnout or reduced effectiveness, which can undermine the advantages of improved health in terms of productivity (Osang & Sarkar, 2008; Agénor, 2015). **Maintaining high health standards in demanding roles can have a negative impact on productivity due to the psychological and physical strain involved.** With a higher level of education, individuals tend to embrace healthier habits and develop a greater awareness of their health. This is believed to have a positive impact on productivity and foster innovation. Nevertheless, the correlation between education and health outcomes varies across different levels of education. Research indicates that pursuing higher education can contribute to advancements in technology and the implementation of effective management strategies. However, it is important to note that **there may be a point where the additional health benefits obtained from education start to yield diminishing returns in terms of productivity**. This phenomenon occurs because individuals with advanced education may already be operating at a high level of efficiency, and additional improvements in health do not result in substantial increases in productivity. On the contrary, the need to uphold rigorous health standards can result in stress and exhaustion, undermining the advantages in productivity (Yu et al., 2022). In addition, the combination of demanding work and the need to prioritise one's health can result in a situation where improvements in health not only reach a plateau but may even lead to decreased productivity because of burnout and fatigue. This holds especially true for individuals with advanced education who may already be operating at maximum efficiency. As their workload increases, the added pressure of taking care of their health can result in burnout, ultimately impacting their overall productivity. The negative interaction effect can be understood because of the diminishing marginal returns of health improvements in the context of individuals with high educational attainment and demanding job roles (Agénor, 2011; Yu et al., 2022).

Furthermore, **there is the substitution effect which suggests that individuals with advanced education tend to prioritise their well-being and work-life balance**

**over earning more money.** As a result, they may choose to work fewer hours or take on fewer responsibilities, which can ultimately lead to a decrease in overall productivity. Individuals with advanced education often enjoy higher incomes, leading them to prioritise a healthy lifestyle and work-life balance over further financial gains. This change in focus may result in a decline in their motivation to work longer hours or assume extra duties, resulting in a decrease in their overall productivity. Striking a balance between work and personal well-being can have a significant impact on productivity, as evidenced by the correlation between health and academic achievement. Research suggests that numerous professionals, especially those with higher incomes, place a greater emphasis on achieving a healthy work-life balance rather than solely focussing on increasing their income. This reflects a larger pattern where personal well-being is given priority over financial gains (Aspen Institute, 2016; FlexJobs, 2023).

It is crucial to consider these interactions holistically rather than in isolation, as examining them separately might underestimate their overall influence on TFP. Enhancing health can boost the effectiveness of education by ensuring that individuals are physically and mentally fit to fully utilise their skills and knowledge. Nevertheless, the negative interaction effect suggests that there might be a threshold beyond which additional enhancements in health do not result in commensurate increases in productivity for individuals with advanced education. The intricate relationship between various factors implies that policies aimed at improving TFP should take a comprehensive approach. It is crucial to strike a balance between investments in healthcare and education to maximise their collective impact (Agénor, 2011; Yu et al., 2022). Therefore, given these insights, we put forth a hypothesis:

**H3: While health and education both positively influence firm-level TFP, their interaction may exhibit negative returns, particularly at higher levels of educational attainment.**

**Data and Methodology**

This study suggests that varying levels of employee education can have specific impacts on improving efficiency, driving technological advancements, and ultimately fostering production growth. In addition, maintaining good health can significantly increase productivity.

Various theories highlight the crucial importance of education in promoting an increase in productivity. Education plays a vital role in shaping the rate of productivity growth. Based on early human capital theories from the 1960s, it was proposed that human capital enhances the quality of labour, leading to an increase in production capacity. Because of this, human capital is a critical factor in the explanation of economic growth (Mankiw, 1995). In neoclassical growth models with exogenous technical advancement, education is seen as an extra factor that boosts productivity. However, education has a crucial impact on production that extends beyond the scope of physical capital. Education plays a crucial role in enhancing productivity through various means such as facilitating the exchange of technology, nurturing innovation, and enhancing the skills and knowledge of the workforce (Nelson & Phelps, 1966). Based on this analysis, it is suggested that the level of education in a country is closely linked to its ability to adopt and effectively utilise new technologies from other nations, thereby contributing to technological progress. Individuals with advanced education have a significant impact on technological advancement through their involvement in research and development of new technologies. Conversely, individuals who have not completed high school may not have the same level of positive impact.

In addition, understanding the link between employees' education and their health is essential for assessing productivity levels. Improved physical health can have a positive impact on work performance and enhance the advantages of formal education, resulting in higher productivity. Nevertheless, the combined effects of these factors can be difficult to separate and analyse (Haider, Kunst, & Wirl, 2021). While it is generally true that better health can lead to increased productivity, the relationship between health and higher education can sometimes have adverse effects. Employees with advanced education and demanding roles may face heightened stress and burnout due to the pressures of maintaining high health standards. This can potentially offset the productivity advantages of improved health (Osang & Sarkar, 2008; Agénor, 2015). As individuals pursue higher levels of education, the potential benefits of improved health may be overshadowed by the mounting stress and pressure they experience. This can ultimately result in decreased productivity among these individuals (Yu et al., 2022). In addition, it is often observed that individuals with advanced education tend to place a higher value on maintaining a healthy lifestyle and achieving a good work-life balance rather than solely focussing on increasing their income. This shift may potentially lead to a decrease in their motivation to work longer hours or assume extra duties, resulting in a decline in overall productivity (Aspen Institute, 2016; FlexJobs, 2023).

Based on these insights, this paper uses fixed effect model as follows to study the correlation of health, education and their interaction impact of firm-level TFP.

$$lnTFP_{i,t} = \varphi_t + \theta_0 + \theta_1 Health_{i,t} + \theta_2 College_{i,t} + \theta_3 HighSchool_{i,t} + \theta_4 Health * College_{i,t} + \theta_5 Health * HighSchool_{i,t} + \varepsilon_{i,t}$$

Given the above explanation, we opt to extract data from Chinese companies listed on the A-share market between the years 2007 and 2022. Subsequently, we put forth the subsequent model. Life expectancy and the educational attainment of employees are used as indices of health and education, respectively.

**Descriptions of Variables**

1. TFP: Referring to Hua Ping (2005), this study uses Interest per Capita to represent a company's Total Factor Productivity. The data is sourced from Wind.

2. Health: Economic theory suggests that human capital in the form of healthy workers contributes to economic growth. Life expectancy is used as a proxy for workforce health, as higher life expectancy generally indicates better health status and lower morbidity. According to the WHO (2002), life expectancy alone is a strong explanatory variable for economic growth. The data on life expectancy is obtained from the National Bureau of Statistics of China.

3. College: Referring to Hua Ping (2005), this study uses the proportion of employees with a college education or higher in A-share listed companies in China from 2007 to 2022 to represent the contribution of higher education. The data is sourced from Wind.

4. HighSchool: Similarly, the proportion of employees with a high school education or lower in A-share listed companies from 2007 to 2022 is used to represent the contribution of non-higher education. The data is sourced from Wind.

**Empirical Results**

|  | lnTFP |
|---|---|
| *Health* | 13.90*** |
|  | (15.45) |
| *College* | 0.203*** |
|  | (3.55) |
| *HighSchool* | -0.0818* |
|  | (-1.65) |
| *Health_HighSchool* | 0.0186 |
|  | (1.63) |
| *Health_College* | -0.0444*** |
|  | (-3.36) |
| *_cons* | -56.21*** |
|  | (-14.43) |
| *N* | 13034 |

**Health and TFP**

Based on the results, it is evident that life expectancy has a significant and positive effect on firm-level TFP. The coefficient for the health variable is 13.90 and is highly significant at the 1% level. This finding provides support for Hypothesis H1, suggesting that 1% increase of employees' life expectancy will lead approximately to 13.9% increase in TFP. The significant coefficient underscores the crucial impact of health on boosting productivity, underscoring the fact that employees who are in good health tend to be more efficient and have fewer absences, resulting in increased overall firm productivity.

From an economic perspective, the substantial influence of health on total factor productivity implies that companies should consider investing in comprehensive health programs. These programs could incorporate regular medical examinations, workplace wellness initiatives, and enhanced healthcare benefits to optimise employee health and productivity. For policymakers, this discovery highlights the significance of investing in public health infrastructure and ensuring access to high-quality healthcare services. These measures are crucial in supporting a robust workforce and ultimately fostering national productivity growth.

**Education and TFP**

The impact of education on TFP is examined by examining the educational attainment of personnel at various levels. The findings reveal varying effects depending on the educational attainment.

Regarding college education, the coefficient is 0.203, which is highly significant at the 1% level. This finding provides strong support for Hypothesis H2.1, suggesting that 1% increase in the percentage of employees with college and above level education can lead to 0.203% increase in firm-level TFP. Individuals who have completed a college education or higher possess the necessary skills to easily embrace new technologies, implement creative strategies, and enhance managerial effectiveness. These factors collectively lead to increased productivity. This discovery emphasises the significance of investing in higher education and ongoing learning opportunities to cultivate a skilled and inventive workforce.

On the other hand, the coefficient for high school education is -0.0818, and it shows some significance at the 10% level. The negative coefficient indicates that lower levels of education may not have a significant impact on productivity improvements, which partially supports Hypothesis H2.2. The skills and knowledge taught in high school may not be enough to achieve significant productivity gains in today's technology-driven work environment.

These findings highlight the significant impact of advanced education on improving productivity. It is important for firms to prioritise and support their employees' pursuit of higher education and continuous professional growth. Policymakers should give top priority to enhancing the quality and availability of higher education to cultivate a highly skilled and efficient workforce that can propel technological progress and foster economic expansion.

**Interaction Effects**

The relationship between health and education offers further understanding of how they impact TFP. The interaction term between health and high school education is positive but statistically insignificant (coefficient = 0.0186, t-value = 1.63). It appears that the impact of health improvements on productivity is not significantly influenced by the lower educational attainment of employees.

On the other hand, there is a significant and negative relationship between health and college education (coefficient = -0.0444, $p < 0.01$). Based on this finding, it seems that there is a point where the benefits of improved health and higher education start to decrease. In other words, once individuals in China reach a certain level of health and education, further improvements may not lead to significant increases in productivity.

These complex interaction effects suggest that the advantages of health and education are influenced by the specific circumstances and can differ depending on the levels of each factor. It is important for firms to carefully consider their investments in health and education to maximise productivity gains, taking into account the possibility of diminishing returns. Policymakers should develop comprehensive policies for human capital development that optimise the synergistic impact of health and education on productivity.

**Conclusion and Discussion**

This study has offered valuable insights into the intricate dynamics between employee education, health, and firm-level TFP in China. Although the research has made significant contributions, it is not without its limitations which need future enhancement. The primary limitation is the use of firm-level data from A-share listed companies, which might not provide a comprehensive representation of the Chinese economy, especially when it comes to smaller firms and other sectors. Future studies could benefit from incorporating a more extensive dataset that encompasses a diverse array of firms and industries, allowing for more robust generalisations of the findings.

In addition, using life expectancy as a measure of health, although informative, does not cover all the factors that affect productivity, such as mental health and workplace wellness. Expanding the scope of health metrics could provide a more profound insight into the connection between health and productivity. In addition, the analysis covers a wide range of data from 2007 to 2022, allowing for a comprehensive understanding of long-term trends. However, it is important to note that this timeframe may not fully capture short-term fluctuations and the potential impacts of recent policies. Future research may consider incorporating more detailed data or including additional control variables to more accurately account for these effects.

The study's results are complicated. Firstly, it is worth noting that life expectancy has a considerable influence on firm-level TFP. Employees who are in good health tend to be more productive, have fewer absences, and work more efficiently. Additionally, pursuing higher education, especially at the college level, has a substantial impact on TFP as it provides individuals with the essential skills needed for technological adaptation, innovation, and effective management. On the other hand, individuals with lower levels of education (high school or below) do not experience significant improvements in productivity. This suggests that the skills acquired at this level are not enough to generate substantial gains in productivity in today's technology-driven work environment.

Furthermore, the interaction between health and education demonstrates that their combined impact can have a negative influence, particularly for employees with a high level of education. This indicates an intricate relationship where further advancements in health might not necessarily lead to a proportional increase in productivity for individuals with advanced education. This could be attributed to factors such as stress or excessive workload in high-pressure job positions.

This study has significant policy and practical implications. The findings highlight the significance of implementing comprehensive human capital policies that improve both health and education simultaneously. Policymakers should develop strategies that maximise the synergistic impact of health and education, taking into account the possibility of diminishing returns. For companies, the findings strongly support the importance of investing in extensive health programs and ongoing education and training for their employees. These investments have the potential to enhance productivity and cultivate a workforce that is more innovative and adaptable.

The study also adds to the theoretical understanding by presenting empirical evidence from firm-level data in China, emphasising the complex and situation-specific relationship between health, education, and productivity. This enhanced understanding can inform more effective policies and business strategies aimed at improving TFP and, consequently, fostering economic growth. Future research directions involve delving into more intricate health metrics, expanding datasets, and conducting in-depth investigations into the mechanisms underlying the diminishing returns of health and education interactions.

This research provides a comprehensive analysis of the effects of health and education on firm-level TFP in China, offering practical insights for policymakers and business leaders. Through addressing the identified limitations and building on the findings, future studies can continue to refine our understanding of these critical factors and how they interact with each other. This will lead to the development of more effective strategies for enhancing productivity and promoting sustainable economic growth.